\documentclass{article}
\usepackage{spconf}
\usepackage{url}
\usepackage{amssymb,amsmath,amsfonts,bm,epsfig,graphicx,theorem,color}

\newcommand{\Eb}{\mathbb{E}}

\newcommand{\lv}{\mathbf{1}}

\newcommand{\Nc}{\mathcal{N}}

\newcommand{\argmin}{\mathop{\mbox{\rm arg\,min}}}

\newcommand{\ssf}[1]{\textrm{$\sf{#1}$}{}}
\newcommand{\norm}[1]{\left \| #1 \right \|}

\newcommand{\Expect}{\mathbb{E}}

\newcommand{\pth}[1]{\left( #1 \right)}

\newcommand{\sth}[1]{\left\{ #1 \right\}}
\newcommand{\iprod}[2]{\left \langle #1, #2 \right\rangle}

\newtheorem{Theorem}{Theorem}

\newtheorem{Lemma}{Lemma}
\newtheorem{Assumption}{Assumption}
\newtheorem{Remark}{Remark}
\newenvironment{Proof}[1]{\medskip\par\noindent
{\bf Proof:\,}\,#1}{{\mbox{\,$\blacksquare$}\par}}

\newcommand{\mypara}[1]{{\smallskip \noindent \bf #1}\hspace{0.1in}}

\graphicspath{{figures/}}

\title{On Federated Learning with Energy Harvesting Clients}
\name{Cong Shen$^*$, Jing Yang$^\dag$, and Jie Xu$^\ddag$ \thanks{CS was supported in part by the National Science Foundation (NSF) under CNS-2002902, ECCS-2029978, and ECCS-2033671. JY was supported in part by the NSF under CNS-1956276, CNS-2003131, ECCS-2030026, and CNS-211454. JX was supported in part by the NSF under ECCS-2033681 and CNS-2044991.}
}
\address{$^*$ University of Virginia\\
$^\dag$The Pennsylvania State University\\ 
$^\ddag$University of Miami}

\begin{document}

\maketitle

\begin{abstract}
Catering to the proliferation of Internet of Things devices and distributed machine learning at the edge, we propose an energy harvesting federated learning (EHFL) framework in this paper. The introduction of EH implies that a client's availability to participate in any FL round cannot be guaranteed, which complicates the theoretical analysis. We derive novel convergence bounds that capture the impact of time-varying device availabilities due to the random EH characteristics of the participating clients, for both parallel and local stochastic gradient descent (SGD) with non-convex loss functions. The results suggest that having a uniform client scheduling that maximizes the minimum number of clients throughout the FL process is desirable, which is further corroborated by the numerical experiments using a real-world FL task and a state-of-the-art EH scheduler. 
\end{abstract}


\section{Introduction}
\label{sec:intro}

Federated learning (FL) is a novel machine learning (ML) paradigm that builds a global ML model by training at many distributed clients. FL represents an ongoing paradigm shift towards moving the data collection and model training away from the server and to the edge \cite{lim2020federated,zhu2020}. The proliferation of Internet of Things (IoT) devices that produce massive amount of data directly at the edge devices, the desire to reduce data transfer to the cloud, and the need to improve ML responsiveness have made FL in IoT networks an important application.

Despite its potential and impact, FL in IoT networks is a difficult task as IoT devices are highly resource constrained. In particular, this paper focuses on enabling FL with energy harvesting (EH) devices \cite{gorlatova2015movers,ejaz2017efficient}, where the computation \cite{guler2021energy,guler2021sustainable,hamdi2021federated} and communication \cite{jing12jcn,Yang:jsac:2015,Yang:jsac:2016,Fong:jsac:2016,Yang_TWC_broadcast} operations of FL at {an EH} device depend entirely on its harvested energy. The focus of FL with EH devices is motivated by the rapid deployment of these devices in IoT networks, such as the agricultural application where devices may be exclusively powered by {ambient energy} sources such as wind or solar \cite{ojha2021internet}. 

The main {challenge}, however, is that the introduction of EH devices complicates the already difficult FL problem. In particular, FL cannot narrowly focus on each learning round, but must consider the temporal correlation of progressive learning rounds that collectively determine the final learning outcome. With EH devices, the availability of any given client is no longer guaranteed for FL in a given round, if it does not have sufficient energy for computation and communication. Furthermore, the random evolution of the energy queue at each device also has temporal correlation that depends on both the energy arrival process and the FL client scheduling algorithm. {The} {coupled temporal correlations} of the FL process and the EH process represent a significant challenge in both theoretical analysis and algorithm design, suggesting that one cannot separately consider the EH design and FL design when optimizing the overall system performance. 

In this paper, we propose an energy harvesting federated learning (EHFL) framework, where EH clients are scheduled to participate in the FL process. To address the aforementioned challenges of EHFL, we first analyze the convergence behavior of FL under an arbitrary sequence of available clients that participate in the corresponding learning rounds. This analysis is useful in that the sequence of clients can be viewed as the output of an EH client scheduler, and optimizing the resulting convergence bound sheds light on the desired behavior of the EH scheduler. A unified principle for both parallel and local stochastic gradient descent (SGD) emerges from the analysis, which suggests that a uniform client scheduling that maximizes the minimum number of clients in FL is beneficial. This theoretical result is corroborated by a numerical experiment using the standard CIFAR-10 classification task and a state-of-the-art EH scheduler.  

\section{The EHFL Framework}
\label{sec:model}

\begin{figure}
    \centering
    \includegraphics[width=0.99\linewidth]{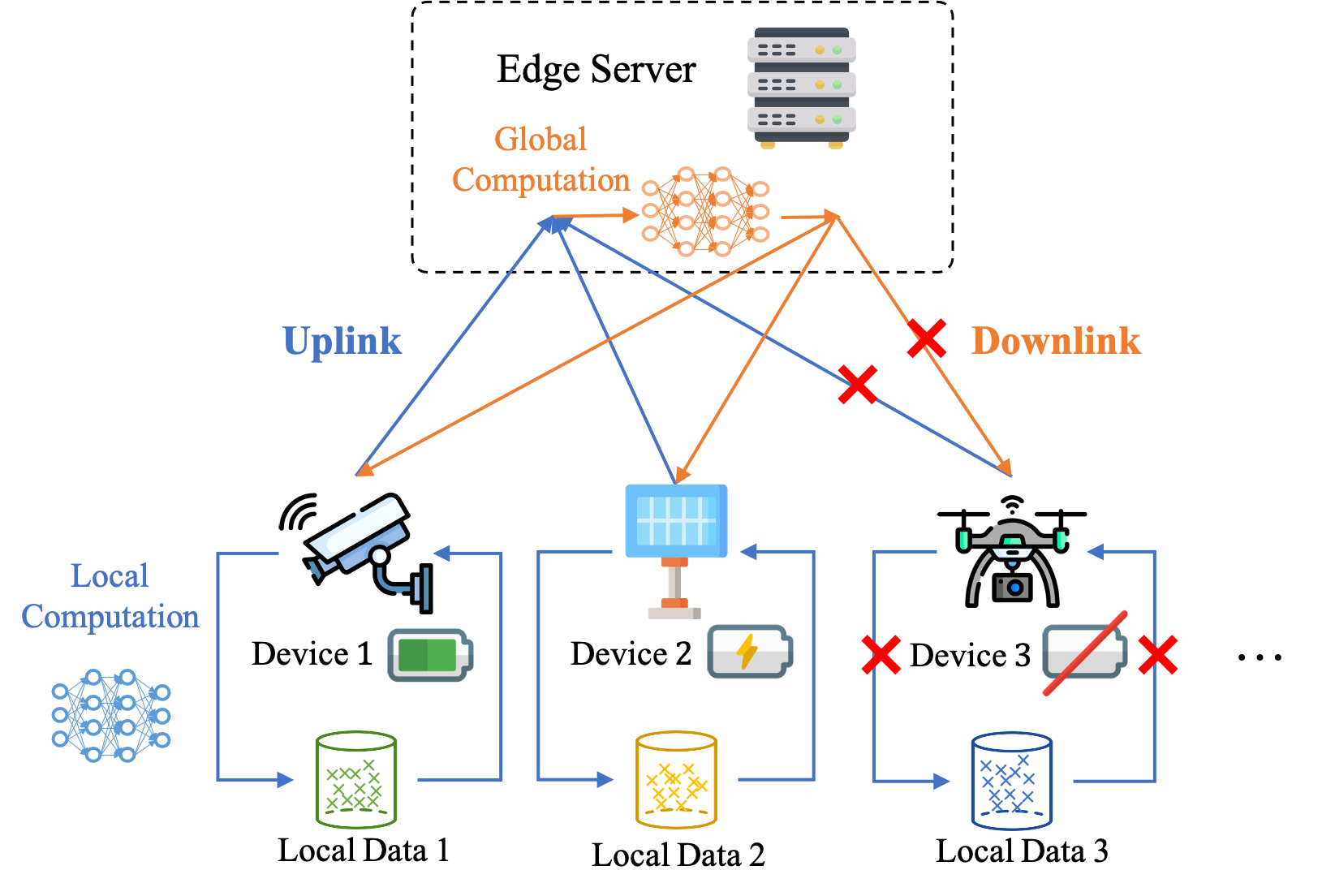}
    \caption{Illustration of the EHFL framework.}
    \label{fig:ehfl}
\end{figure}

The proposed energy harvesting federated learning (EHFL) framework is illustrated in Fig.~\ref{fig:ehfl}. This framework is notably different from standard FL, because the introduction of EH devices implies that a client's availability to participate in any round cannot be guaranteed. FL must deal with different sets of available clients that are determined \emph{exogenously} (by the EH scheduler) in every round, which would affect the model convergence. To further complicate the analysis, such client availability is not independent over time, as clients who {have} participated in one round and consumed the harvested energy are less likely to have sufficient energy for the next round. 

\mypara{Federated learning model.}  In a typical case, the goal of FL is to solve the standard empirical risk minimization (ERM) problem:
 \begin{equation*}
\min_{x\in \mathbb{R}^d} f(x) = \min_{x\in \mathbb{R}^d} \frac{1}{D} \sum_{z \in \mathcal{D}} l(x; z),
 \end{equation*}
in a distributed fashion, where $x\in\mathbb{R}^d$ is the machine learning model variable that we would like to optimize, $l(x; z)$ is the loss function evaluated at model $x$ and data sample $z$, and $f: \mathbb{R}^d\rightarrow\mathbb{R}$ is the differentiable loss function averaged over the total dataset $\mathcal{D}$ with size $D$. 
We denote $x^* := \argmin_{x \in \mathbb{R}^d} f(x)$, and $f^* := f(x^*)$. We denote the \emph{maximum} number of clients in the FL system as $M$, and the total global dataset is the union of all local datasets at these $M$ clients: $\mathcal{D} = \bigcup_{m=1}^M \mathcal{D}_m$. We assume that $\mathcal{D}_i$ has $D_i$ data samples at client $i \in [M] :=\sth{1,\cdots,M}$, and all local datasets are non-overlapping, hence $\sum_{i}D_i=D$. Note that $M$ is generally not the number of clients that participate in FL in any given learning round. 
The original ERM problem can be rewritten as   
 \begin{equation*}
\min_{x\in \mathbb{R}^d} f(x)=\min_{x\in \mathbb{R}^d}\sum_{i=1}^{M} \frac{D_i}{D} f_i(x),
 \end{equation*}
where  $f_i: \mathbb{R}^d\rightarrow\mathbb{R}$ is the local loss function for client $i$, averaged over its local dataset $\mathcal{D}_i$, i.e., $f_i(x) = \frac{1}{D_i} \sum_{\xi \in \mathcal{D}_i} l(x; \xi)$. 

We consider that local SGD \cite{stich2018local} is adopted to solve the FL problem. In the $t$-th round of local SGD, $t=1, \cdots, T$, there are $n_t$ clients $\Nc_t := \{m_1, \cdots, m_{n_t}\}$ who actively participate in FL. Each client independently runs {$K$} individual SGD steps before aggregating the local models at the server. Specifically, the $t$-th round starts with client $i \in \Nc_t$ receiving the latest global model $x_t$ from the parameter server: $x_t^i = x_{t,0}^i = x_t$. It then runs {$K$} steps of stochastic gradient evaluation:  
 \begin{equation} \label{eqn:locSGDupdt}
x_{t, \tau+1}^i = x_{t, \tau}^i - \eta_t \nabla \tilde{f}_i(x_{t, \tau}^i ), \forall \tau=0, \cdots, {K-1}.
 \end{equation}
The client's updated model after these $K$ steps can be written as $x_{t+1}^i = x_{t, {K}}^i$. 
Notation wise, we use $\tilde{f}_i(x) := l(x; \xi_i)$ to denote the loss function of model $x$ evaluated with a random data sample $\xi_i$ at client $i$. The server collects the local models $\{x_{t+1}^i, i \in \Nc_t\}$ and computes a simple aggregation $x_{t+1} = \frac{1}{n_t} \sum_{i \in \Nc_t } x_{t+1}^i$ as the global model for the next round. Local SGD then moves on to the $(t+1)$-th round.

\mypara{Energy harvesting model.} 
In EHFL, each client $i \in [M]$ is powered by energy harvested from the ambient environment. We assume that each client has an energy queue (rechargeable batteries or capacitors) to store the harvested energy. The energy queue at each client is replenished randomly and consumed by computation and communication for FL. 
We assume that the energy unit is normalized so that if a device participates in one round of FL, it consumes one unit of energy. This energy unit represents the cost of both computation and communication. We assume the duration between two consecutive rounds is fixed. 

Let $E_i(t)$ be the total amount of energy units available at the beginning of round $t$ at device $i$, and $A_i(t)$ be the amount of energy units harvested during the $t$-th round. We assume $A_i(t)$ is an independent and identically distributed (IID) {Bernoulli} random variable with $\Eb[A_i(t)]=\lambda_i$. Different values of $\lambda_i$ capture the energy heterogeneity among clients.  Then, the energy level at device $i$ evolves according to the following equation:
\begin{equation}\label{eqn:ener_queue}
    E_i(t+1)=\min\{\pth{E_i(t)-\lv\{i\in\Nc_t\}}+A_i(t), E_{\max}\}
\end{equation}
where $\lv\{\cdot\}$ is the indicator function, $E_{\max}$ is the capacity of the battery, and the energy causality condition requires that $E_i(t)\geq \lv\{i\in\Nc_t\}$ for all $i,t$.

\section{Convergence analysis for EHFL}
\label{sec:analysis}

We analyze the convergence of FL with an arbitrary sequence of participating clients $\{\Nc_1,  \cdots, \Nc_T\}$ as the output of the EH scheduler, with non-convex loss functions. We first focus on a special case of \emph{parallel SGD}, which refers to distributed SGD with per-step model average, to gain some insight of the FL convergence behavior due to the random EH characteristics. We  then extend the analysis to \emph{local SGD} with periodic model average whose period is strictly larger than one. Finally we summarize the main theoretical result and discuss its implication on the EH scheduler design.

\subsection{Parallel SGD: $K=1$}
\label{sec:analysis_psgd}

\subsubsection{Assumptions}
We limit our attention to $L$-smooth (possibly non-convex) loss functions, as stated in Assumption \ref{assu_Lsmo}.  In addition, we assume that the stochastic gradients are unbiased at all clients, and the variance is (uniformly) bounded in Assumption \ref{assu_sgd}. 

\begin{Assumption}
\label{assu_Lsmo}
$l(x,\xi)$ is $L$-smooth:  $\| \nabla l(x,\xi)-\nabla l(y,\xi) \|\leq L \| x-y \| $ for any $x,y\in \mathbb{R}^d$ and any $\xi \in \mathcal{D}$. 
\end{Assumption}

\begin{Assumption}
\label{assu_sgd}
SGD is unbiased at all clients: $\mathbb{E}_{\xi} \nabla f_i(x) =  \nabla f(x), \forall i$, and its variance is bounded: $ \mathbb{E}_{\xi} \|\mathbb{}  \nabla l(x, \xi) - \nabla f(x)\|^2 \leq \sigma^2 $.
\end{Assumption}

\subsubsection{Main result}

We note that for non-convex loss functions, it is well-known that SGD may converge to a local minimum or saddle point, and it is a common practice to evaluate the expected gradient norms as an indicator of convergence. In particular, an algorithm achieves an $\epsilon$-suboptimal solution if 
$\frac{1}{T} \sum_{t=0}^{T-1}\mathbb{E} \|\nabla f(x_{t}) \|^2 \leq \epsilon$, which guarantees the convergence to a stationary point \cite{wang2018cooperative}.

We now state our main result in Theorem~\ref{thm_main_psgd}. Detailed proofs of both theorems can be found in the  Appendix. 

\begin{Theorem}
\label{thm_main_psgd}
Suppose Assumptions \ref{assu_Lsmo} and \ref{assu_sgd} hold. Consider an energy harvesting client scheduler that produces $n_t$ clients to participate in the $t$-th round parallel SGD. Assume $0< n_{\min} \leq n_t \leq n_{\max} \leq M$, and we choose a parameter $\eta$ satisfying $0< \eta \leq  \frac{1}{L} \sqrt{ \frac{T}{n_{\max}} } $. Then, if we set the learning rate of SGD as $$\eta_t = \eta \sqrt{\frac{n_t}{T}}, , \forall t=0, \cdots, T-1,$$ the convergence of parallel SGD with non-convex loss functions and IID local datasets satisfies: 
\begin{align}   \label{eqn:nc_convg}
  & \frac{1}{T}  \sum_{t=0}^{T-1} \mathbb{E}\|\nabla f(x_{t}) \|^2 \leq \frac{f(x_{0}) -f^*}{ \eta \sqrt{ n_{\min} T} - \frac{L}{2} \eta^2 n_{\min} }  \nonumber \\
  & + \frac{L \sigma^2}{2 \eta \sqrt{n_{\min} T} - L\eta^2 n_{\min}  } \sim \mathcal{O} \pth{\frac{1}{\sqrt{ n_{\min} T }}}.
\end{align}
\end{Theorem}

\begin{Remark}
\normalfont
The key novelty in this theorem is to establish the relationship $\eta_t = \eta \sqrt{\frac{n_t}{T}}$, which is accomplished by minimizing the derived upper bound as a general function of $\eta_t$ and $n_t$.  
Theorem~\ref{thm_main_psgd} states that if we tie the choice of learning rate to the available number of clients according to $\eta_t \sim \mathcal{O}\pth{\sqrt{n_t}} $, then we achieve the same $\mathcal{O} \pth{{1}/{\sqrt{T}}}$ convergence rate as the constant-client parallel SGD \cite{stich2018local}.
\end{Remark}

\begin{Remark}
\normalfont
It is known that within a proper range that guarantees the convergence, selecting larger {stepsize} has the benefit of speeding up the SGD process.  In this spirit, a particular choice of $\eta$ is $\eta = \frac{1}{L} \sqrt{ \frac{T}{n_{\max}} }$, which leads to $\eta_{\min} := \min{\eta_t} = \frac{1}{L} \sqrt{\frac{n_{\min}}{n_{\max}}}$. This results in a convergence scaling of $\mathcal{O} \pth{ \sqrt{\frac{n_{\max}}{n_{\min}}}  \frac{1}{\sqrt{T} }   }$.  Clearly, selecting a uniform client scheduling such that $n_{\max}=n_{\min}$ minimizes the coefficient of $ \frac{1}{\sqrt{T} }$. This insight thus provides a theoretical guidance for the EH scheduler design.
\end{Remark}

\begin{Remark}
\normalfont
Assumption \ref{assu_sgd} corresponds to the so-called IID local dataset setting for FL. How to extend the analysis to non-IID local datasets is an interesting future research direction.
\end{Remark}

\subsection{Local SGD: $K >1$}
\label{sec:analysis_lsgd}

We now analyze the case of local SGD with $K>1$. The main result is stated as follows.

\begin{Theorem}
\label{thm_main_lsgd}
Suppose Assumptions \ref{assu_Lsmo} and \ref{assu_sgd} hold. Consider an energy harvesting client scheduler that produces $n_t$ clients to participate in the $t$-th round local SGD. Assume $0< n_{\min} \leq n_t \leq n_{\max} \leq M$, and we choose a parameter $\eta$ satisfying $0< \eta \leq  \frac{1}{2KL} \sqrt{ \frac{1}{30 n_{\max}} } $. Then, if we set the stepsize of SGD at the $t$-th round as $$ \eta_t = \eta \sqrt{\frac{n_t}{T}}, \forall t=0, \cdots, T-1,$$ then we achieve the following convergence of local SGD with non-convex loss functions:
\begin{align}  
& \frac{1 }{T}  \sum_{t=0}^{T-1} \mathbb{E}\|\nabla f(x_{t}) \|^2 \leq 
 \frac{ \frac{2}{K}  \pth{f(x_{0}) -f^*} + L \sigma^2 \eta^2 }{  \eta \sqrt{n_{\min}T} - \sqrt{30} KL \eta^2 n_{\min}} \nonumber \\
& + \frac{5KL^2 \sigma^2 \eta^3 n_{\max}^{\frac{3}{2}} }{  \eta \sqrt{n_{\min} T} - \sqrt{30} KL \eta^2  n_{\min}  \sqrt{T} } = \mathcal{O} \pth{\frac{1}{ \sqrt{n_{\min} T} }}. \label{eqn:nc_convg_lsgd} 
\end{align}
\end{Theorem}

\begin{Remark}
\normalfont
The key challenge for analyzing local SGD is that the gradient estimation after the first step becomes \emph{biased}, i.e., they do not represent the true gradients in expectation. Having a varying $n_t$ means that different rounds are ``heterogeneous'' in terms of averaging the biased SGDs with varying variances, which cannot be easily handled when bounding the convergence rate. The proof relies on enhancing the \emph{perturbed iterate framework} \cite{mania2017siam} to decouple the impact of each additional SGD step by a careful construction of the virtual model sequence. This allows us to derive an $\eta_t$-dependent upper bound for the average (over $n_t$ clients) gradient for each SGD step $\tau=0, \cdots, {K-1}$. This bound is then {utilized} in the enhanced perturbed iterate framework to derive a non-trivial $(n_t, \eta_t)$-dependent convergence rate upper bound. Then, similar to Theorem~\ref{thm_main_psgd}, we can minimize this bound over the choice of $\eta_t$ as a function of $n_t$. 
\end{Remark}

\begin{Remark}
\normalfont
Theorem~\ref{thm_main_lsgd} unifies the selection of learning rate as a function of the EH device availability for both parallel and local SGDs (at least with respect to the scaling), which suggests that the EH scheduler design can be agnostic to the SGD steps chosen by the FL task. This is an important feature that improves the generalization of the proposed EHFL framework in terms of the performance guarantees.
\end{Remark}

\subsection{EH scheduler design}
\label{sec:EHsche}

The convergence analysis for both parallel and local SGD indicates that maintaining a balanced number of clients participating in each round throughout the learning horizon is desirable. However, strictly maintaining a constant number of clients in the face of stochastic energy arrival and energy causality constraint is a very challenging task, not to mention the inhomogeneous EH processes at clients. 

In order to gain some intuition of the desired EH scheduler design, we first ignore the stochasticity of the EH process and focus on the long-term average EH rate instead.
Given the total EH rate $\Lambda:=\sum_{i=1}^M\lambda_i$ and the energy flow conservation condition (i.e., energy consumption rate must be upper bounded by the energy arrival rate), the average number of active clients in each round must be upper bounded by $\Lambda$ as well. For a clear exposition of our rationale, we assume $\Lambda$ is an integer. Thus, if we are able to obtain a subset of clients $\Nc_t$ in round $t$ such that $|\Nc_t|=\Lambda$ with high probability, then we can expect that the $n_{\min}$ throughout the learning process is maximized, and the convergence rate can thus be optimized with high probability based on our theoretical results. The problem then boils down to ensuring such a selection of $\Nc_t$ is feasible in each round, in the presence of stochastic energy arrivals and heterogeneous EH rates across the clients. 

In our previous work~\cite{Yang:jsac:2015}, we have developed an energy queue length based {\it myopic} scheduling policy when $E_{\max}=\infty$. At the beginning of round $t$, the scheduler first selects $\Lambda$ clients with the longest energy queues and forms a candidate set of active clients, denoted as $\Nc_t'$. Then, it determines  $\Nc_t=\{i: i\in \Nc_t', E_i(t)\geq 1\}$. 
The myopic scheduling policy has a queue-length balancing nature, i.e., it tries to equalize the battery levels of all clients by prioritizing clients with longer energy queues. {As a result, it ensures that $|\Nc_t|=\Lambda$ in almost every round $t$.} We will evaluate the performance of this myopic EH scheduling policy in the experiment.

\section{Simulation Results}
\label{sec:sim}

\begin{figure}
    \centering
    \includegraphics[width=0.99\linewidth]{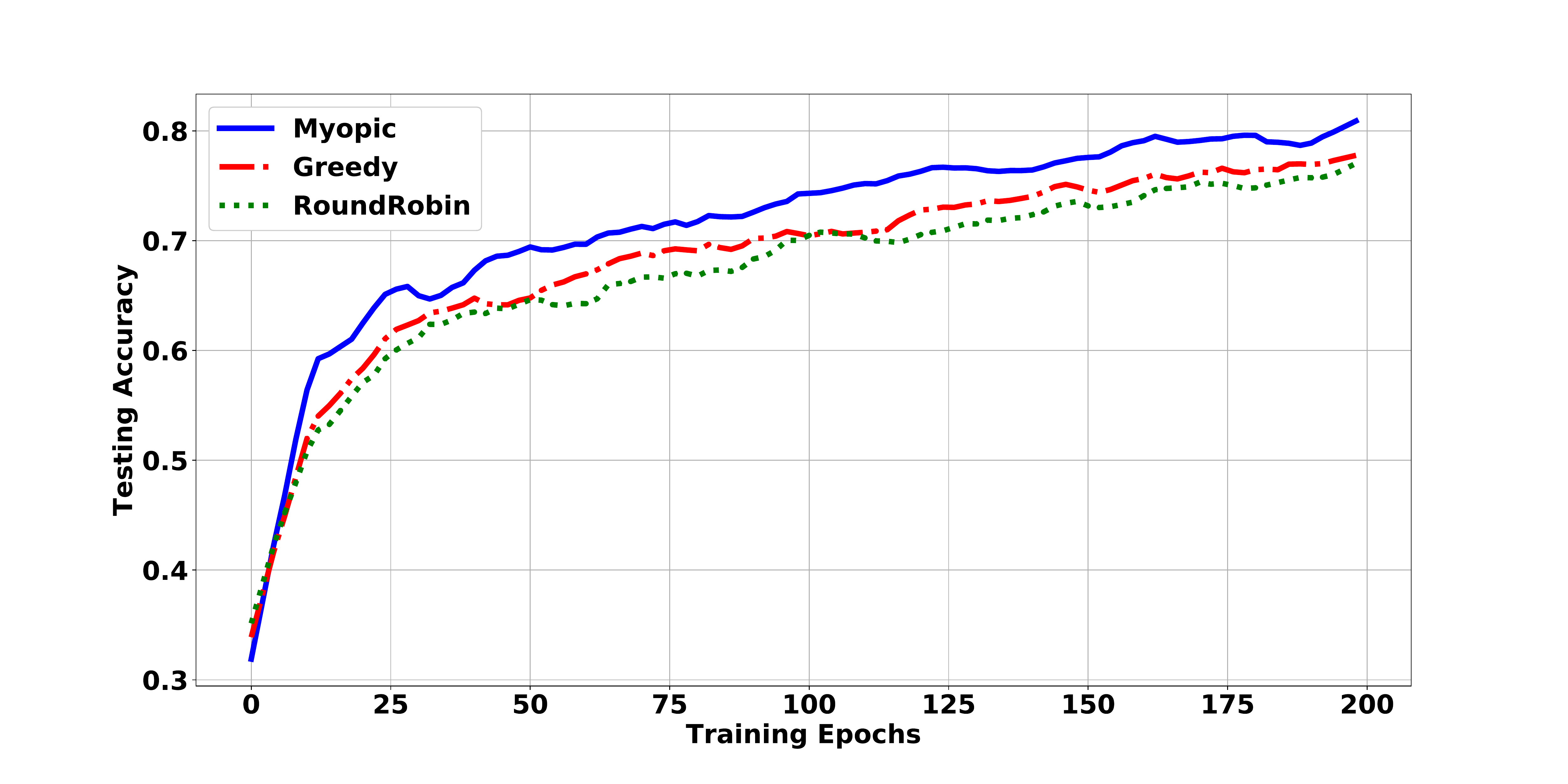}
    \caption{Model convergence comparison of \emph{Myopic} \cite{Yang:jsac:2015} with two baseline EH schedulers \emph{Round Robin} and \emph{Greedy} for EHFL.}
    \label{fig:myopic}
\end{figure}

\mypara{Experiment setup.} We have carried out an experiment on the standard real-world CIFAR-10 classification task \cite{krizhevsky2009learning} under the proposed EHFL framework. We set $M=10$, $K=5$, and mini-batch size of 50. The nominal learning rate initially sets to $0.15$ and decays every 10 rounds with rate 0.99. On top of that, we apply a $c\sqrt{n_t}$ variation such that the mean value for every 10 rounds remain the same as the nominal learning rate. We train a convolutional neural network (CNN) model with two $5 \times 5$ convolution layers (both with 64 channels), two fully connected layers (384 and 192 units respectively) with $\ssf{ReLU}$ activation and a final output layer with softmax. The two convolution layers are both followed by $2 \times 2$ max pooling and a local response norm layer. In each round, the available clients are generated by the corresponding EH scheduler, and will participate in FL if its available energy is larger than one unit. Otherwise, the client will not participate in FL in the current round. We set $\Lambda=5$ with a homogeneous arrival rate of all clients for the \emph{Myopic} policy of \cite{Yang:jsac:2015}.

\mypara{Main result.}  The model convergence performances of EHFL under three EH schedulers are plotted in Fig.~\ref{fig:myopic}. The \emph{Round Robin} policy cyclically schedule among all clients, while the \emph{Greedy} policy always schedule the clients with {non-empty energy queues}. We can see that the \emph{Myopic} policy has the best performance among the three scheduler, while  \emph{Round Robin} has the worst convergence.

\section{Conclusions}
We have carried out a novel convergence analysis of federated learning under an arbitrary sequence of participating clients for each learning round, for non-convex loss functions and both parallel and local SGD. The analysis revealed a unified client scheduling principle, which is to maintain a balanced number of clients participating in each round throughout the learning horizon. This result offers a principled guideline for the energy harvesting client scheduler design, and we have shown via a numerical experiment that a state-of-the-art energy harvesting scheduler that follows this guideline achieves better convergence performance for a standard real-world FL task.

\bibliographystyle{IEEEbib}
\bibliography{FedLearn,shen,yang,xu}

\newpage
\onecolumn
\appendix

\section{Proof of Theorem~1}

\begin{Proof}
The server model update at the end of round $t$ is
\begin{equation*}
x_{t+1} = \frac{1}{n_t} \sum_{i \in \Nc_t}  x_{t+1}^i = \frac{1}{n_t} \sum_{i \in \Nc_t}  \pth{ x_t - \eta_t  \nabla f_i(x_{t}) } = x_t - \frac{\eta_t}{n_t} \sum_{i \in \Nc_t}   \nabla f_i(x_{t}).
\end{equation*}
We can evaluate the average loss with respect to model $x_{t+1}$ as
\begin{eqnarray} \label{eqn:thm_main_psgd_1}
\Expect{f(x_{t+1})} &=& \Expect  f\pth{x_t - \frac{\eta_t}{n_t} \sum_{i \in \Nc_t}   \nabla f_i(x_{t}) } \nonumber \\
& \leq& \Expect{f(x_t)} - \frac{\eta_t }{n_t}\Expect{  \iprod{ \nabla f(x_{t})}{ \sum_{i \in \Nc_t} \nabla f_i(x_{t})} }  + \frac{L\eta_t^2}{2} \Expect{ \norm{ \frac{1}{n_t} \sum_{i \in \Nc_t} \nabla f_i(x_{t}) }^2} \nonumber \\
& =& \Expect{f(x_t)} - \eta_t \Expect{ \norm{\nabla f(x_{t})}^2 } + \frac{L\eta_t^2}{2} \Expect{ \norm{ \frac{1}{n_t} \sum_{i \in \Nc_t} \nabla f_i(x_{t}) }^2}. 
\end{eqnarray}
We analyze the last term in Eqn.~\eqref{eqn:thm_main_psgd_1}, and have
\begin{eqnarray} \label{eqn:thm_main_psgd_2}
\Expect{ \norm{ \frac{1}{n_t} \sum_{i \in \Nc_t} \nabla f_i(x_{t}) }^2} &=&\Expect{ \norm{\frac{1}{n_t} \sum_{i \in \Nc_t}  \nabla f_i(x_{t})  -  \nabla f(x_{t}) + \nabla f(x_{t})}^2}  \nonumber \\
&=& \Expect_{x_t} \left[ \Expect_{\xi} \norm{\frac{1}{n_t} \sum_{i \in \Nc_t}  \nabla f_i(x_{t})  -  \nabla f(x_{t}) + \nabla f(x_{t})}^2  | x_t \right] \nonumber \\
&=& \Expect_{x_t} \left[ \Expect_{\xi} \norm{\frac{1}{n_t}\sum_{i \in \Nc_t}  \left( \nabla f_i(x_{t})  -  \nabla f(x_{t}) \right) }^2 + \norm{\nabla f(x_{t})}^2  | x_t \right] \nonumber \\
&\leq& \frac{\sigma^2}{n_t}  + \Expect{ \norm{\nabla f(x_{t})}^2 }.
\end{eqnarray}
Plugging Eqn.~\eqref{eqn:thm_main_psgd_2} back to \eqref{eqn:thm_main_psgd_1} leads to
\begin{eqnarray} \label{eqn:thm_main_psgd_3}
\Expect{f(x_{t+1})}  &\leq& \Expect{f(x_t)} - \eta_t \Expect{ \norm{\nabla f(x_{t})}^2 } + \frac{L\eta_t^2}{2} \Expect{ \norm{ \frac{1}{n_t} \sum_{i \in \Nc_t} \nabla f_i(x_{t}) }^2} \nonumber \\
&\leq&  \Expect{f(x_t)} - \eta_t \Expect{ \norm{\nabla f(x_{t})}^2 } + \frac{L\eta_t^2}{2} \Expect{ \norm{\nabla f(x_{t})}^2 } + \frac{L\eta_t^2 \sigma^2}{2n_t} \nonumber \\
&=&   \Expect{f(x_t)} - \pth{\eta_t - \frac{L\eta_t^2}{2}} \Expect{ \norm{\nabla f(x_{t})}^2 } + \frac{L\eta_t^2 \sigma^2}{2n_t}.
\end{eqnarray}
Eqn.~\eqref{eqn:thm_main_psgd_3} is equivalent to
\begin{equation} \label{eqn:thm_main_psgd_4}
\pth{\eta_t - \frac{L\eta_t^2}{2}}  \Expect{ \norm{\nabla f(x_{t})}^2 } \leq \Expect{f(x_t)} - \Expect{f(x_{t+1})} + \frac{Ln_t^2 \sigma^2}{2n_t},
\end{equation}
and we can further sum Eqn.~\eqref{eqn:thm_main_psgd_4} from 0 to $T-1$ and average, resulting in
\begin{eqnarray} \label{eqn:thm_main_psgd_5}
 \frac{1}{T} \sum_{t=0}^{T-1} \pth{\eta_t - \frac{L\eta_t^2}{2}}  \Expect{ \norm{\nabla f(x_{t})}^2 } &\leq&  \frac{1}{T} \pth{ f(x_0) - \Expect{f(x_T)}  }  +  \frac{1}{T} \sum_{t=0}^{T-1}  \frac{L \eta_t^2 \sigma^2}{2n_t} \nonumber \\
 &\leq& \frac{1}{T} \pth{f(x_0) - f^* } + \frac{L\sigma^2}{2T} \sum_{t=0}^{T-1} \frac{\eta_t^2}{n_t}.
\end{eqnarray}
The condition $0< \eta \leq  \frac{1}{L} \sqrt{ \frac{T}{n_{\max}} }$ implies the function  $\eta_t - \frac{L\eta_t^2}{2}$ is both positive and monotonically increasing with $\eta_t$. Hence 
\begin{eqnarray} \label{eqn:thm_main_psgd_6}
 \frac{1}{T} \sum_{t=0}^{T-1} \pth{\eta_{\min} - \frac{L\eta_{\min}^2}{2}}  \Expect{ \norm{\nabla f(x_{t})}^2 }   &\leq& \frac{1}{T} \sum_{t=0}^{T-1} \pth{\eta_t - \frac{L\eta_t^2}{2}}  \Expect{ \norm{\nabla f(x_{t})}^2 } \nonumber \\
 &\leq& \frac{1}{T} \pth{f(x_0) - f^* } + \frac{L\sigma^2 \eta^2}{2T} 
\end{eqnarray}
where \eqref{eqn:thm_main_psgd_6} comes from plugging $\eta_t = \eta \sqrt{\frac{n_t}{T}}$  in \eqref{eqn:thm_main_psgd_5}.  Dividing the $t$-independent $\pth{\eta_{\min} - \frac{L\eta_{\min}^2}{2}}$ from both sides of Eqn.~\eqref{eqn:thm_main_psgd_6}  and plugging in $\eta_{\min} =  \eta \sqrt{\frac{n_{\min}}{T}}$ complete the proof.
\end{Proof}

\section{Proof of Theorem~2}

\begin{Proof}

Some preparation is necessary to facilitate this proof. First of all, the following lemma from \cite[Lemma 3]{reddi2020adaptive} is useful.

\begin{Lemma} \label{lem:reddi}
For $\eta_t \leq 1/\pth{8KL}$ we have
\begin{equation*}
 \frac{1}{n_t} \sum_{i \in \Nc_t} \Expect \norm{ x_t - x_{t,\tau}^i  }^2 \leq 5K \sigma^2 \eta_t^2 + 30 K^2 \eta_t^2 \norm{  \nabla f(x_{t})}^2.
\end{equation*}
\end{Lemma}

Next we define some new variables to simplify the derivation. We denote 
\begin{eqnarray*}
\nabla g_t^i &:=& \sum_{\tau=0}^{K-1} \nabla f_i\pth{x_{t,\tau}^i} \\
\nabla g_t &:=& \frac{1}{n_t} \sum_{i \in \Nc_t} \nabla g_t^i =  \frac{1}{n_t} \sum_{i \in \Nc_t} \sum_{\tau=0}^{K-1} \nabla f_i\pth{x_{t,\tau}^i}  \\
\nabla \bar{g}_t &:=& \Expect_{\{\xi_i\}}{\nabla g_t } = \frac{1}{n_t} \sum_{i \in \Nc_t} \sum_{\tau=0}^{K-1} \nabla f\pth{x_{t,\tau}^i}
\end{eqnarray*}

We start with
\begin{eqnarray} \label{eqn:lsgd_1}
f(x_{t+1}) &=& f\pth{ \frac{1}{n_t} \sum_{i \in \Nc_t} x_{t+1}^i } \nonumber \\
&=& f \pth{  x_t - \eta_t  \nabla g_t } \nonumber \\
&\leq& f(x_t) - \iprod{ \nabla f(x_{t}) }{ \eta_t \nabla g_t } + \frac{L}{2} \eta_t^2 \norm{ \nabla g_t }^2 \nonumber \\
&=& f(x_t) -\eta_t  K \norm{\nabla f(x_{t})}^2 + \eta_t \iprod{ \nabla f(x_{t}) }{ K \nabla f(x_{t}) - \nabla g_t }  + \frac{L \eta_t^2}{2} \norm{ \nabla g_t }^2.
\end{eqnarray}
The next steps are to separately analyze the expectation of the last two terms in Eqn.~\eqref{eqn:lsgd_1}. We first have
\begin{eqnarray} \label{eqn:lsgd_2}
&& \Expect \iprod{ \nabla f(x_{t}) }{ K \nabla f(x_{t}) - \nabla g_t } = \Expect \iprod{ \nabla f(x_{t}) }{ \frac{1}{n_t} \sum_{i \in \Nc_t} \sum_{\tau=0}^{K-1} \pth{ \nabla f(x_{t}) - \nabla f(x_{t,\tau}^i)}  } \nonumber \\
&\overset{(b1)}{=}& \frac{K}{2} \Expect \norm{ \nabla f(x_{t}) }^2  + \frac{1}{2K n_t^2} \Expect \norm{  \sum_{i \in \Nc_t} \sum_{\tau=0}^{K-1} \pth{   \nabla f(x_{t}) - \nabla f(x_{t,\tau}^i)}   }^2  - \frac{1}{2K} \Expect \norm{  \nabla \bar{g}_t  }^2  \nonumber \\
&\overset{(b2)}{\leq}& \frac{K}{2} \Expect \norm{ \nabla f(x_{t}) }^2  + \frac{1}{2n_t} \sum_{i \in \Nc_t} \sum_{\tau=0}^{K-1} \Expect \norm{  \nabla f(x_{t}) - \nabla f(x_{t,\tau}^i)   }^2   - \frac{1}{2K} \Expect \norm{  \nabla \bar{g}_t  }^2   \nonumber \\
&\leq& \frac{K}{2} \Expect \norm{ \nabla f(x_{t}) }^2  +\frac{L^2}{2n_t} \sum_{i \in \Nc_t} \sum_{\tau=0}^{K-1} \Expect \norm{  x_{t} - x_{t,\tau}^i   }^2   - \frac{1}{2K} \Expect \norm{  \nabla \bar{g}_t  }^2   \nonumber \\ 
&\overset{(b3)}{\leq}& \frac{K}{2} \pth{  1+30K^2L^2\eta_t^2 } \Expect \norm{ \nabla f(x_{t}) }^2 + \frac{5K^2L^2\sigma^2\eta_t^2}{2} - \frac{1}{2K} \Expect \norm{  \nabla \bar{g}_t  }^2 
\end{eqnarray} 
where (b1) is because  $\iprod{x}{y} =\frac{1}{2} \norm{x}^2 + \frac{1}{2} \norm{y}^2 -\frac{1}{2} \norm{x-y}^2$, (b2) is due to Cauchy-Schwartz, and (b3) is from Lemma~\ref{lem:reddi}. 

We then evaluate the expectation of the last term of Eqn.~\eqref{eqn:lsgd_1}.
\begin{eqnarray} \label{eqn:lsgd_3}
\Expect \norm{ \nabla g_t }^2 &=& \Expect \norm{  \frac{1}{n_t} \sum_{i \in \Nc_t} \sum_{\tau=0}^{K-1} \nabla f_i(x_{t,\tau}^i)  }^2 \nonumber \\
&\overset{(b4)}{=}& \frac{1}{n_t} \Expect \norm{  \sum_{i \in \Nc_t} \sum_{\tau=0}^{K-1}  \pth{ \nabla f(x_{t,\tau}^i)  - \nabla f_i(x_{t,\tau}^i)   }  }^2 +  \Expect \norm{  \nabla \bar{g}_t  }^2  \nonumber \\ 
&\leq& \frac{K\sigma^2}{n_t} + \Expect \norm{  \nabla \bar{g}_t  }^2
\end{eqnarray} 
where (b4) uses the fact that the SGD sampling error is independent of other random variables. 

Putting both Eqns.~\eqref{eqn:lsgd_2} and \eqref{eqn:lsgd_3} back to the expectation of Eqn.~\eqref{eqn:lsgd_1}, we have
\begin{eqnarray} \label{eqn:lsgd_4}
\Expect f(x_{t+1}) & \leq & \Expect f(x_{t}) - \frac{K}{2} \pth{\eta_t - 30K^2L^2\eta_t^3} \Expect \norm{ \nabla f(x_{t}) }^2 + \frac{5K^2L^2\sigma^2\eta_t^3}{2} + \frac{KL\sigma^2 \eta_t^2}{2n_t}  + \pth{ \frac{L\eta_t^2}{2} - \frac{\eta_t}{2K}  } \Expect \norm{  \nabla \bar{g}_t  }^2  \nonumber \\
&\overset{(b5)}{\leq}& \Expect f(x_{t}) -\frac{K}{2} \pth{\eta_t - 30K^2L^2\eta_t^3} \Expect \norm{ \nabla f(x_{t}) }^2 + \frac{5K^2L^2\sigma^2\eta_t^3}{2} + \frac{KL\sigma^2 \eta_t^2}{2n_t}
\end{eqnarray} 
where (b5) is because for the choice of $\eta \leq  \frac{1}{2KL} \sqrt{ \frac{1}{30 n_{\max}} } $ we can guarantee $\eta_t \leq 1/\pth{2 \sqrt{30}KL} < 1/\pth{KL}$, and thus $$   \frac{L\eta_t^2}{2} - \frac{\eta_t}{2K} \leq 0.$$ Now, rearranging terms of both sides in Eqn.~\eqref{eqn:lsgd_4} and averaging over $t=0$ to $t=T-1$ leads to
\begin{eqnarray} \label{eqn:lsgd_5}
\frac{1}{T} \sum_{t=0}^{T-1} \frac{K}{2} \pth{\eta_t - 30K^2L^2\eta_t^3} \Expect \norm{ \nabla f(x_{t}) }^2 &\leq& \frac{f(x_0) - \Expect f(x_T)}{T} + \frac{1}{T} \sum_{t=0}^{T-1} \frac{KL\sigma^2 \eta_t^2}{2n_t} + \frac{1}{T} \sum_{t=0}^{T-1} \frac{5K^2L^2\sigma^2\eta_t^3}{2} \nonumber \\
&\leq& \frac{f(x_0) - f^*}{T} + \frac{1}{T} \sum_{t=0}^{T-1} \frac{KL\sigma^2 \eta_t^2}{2n_t} + \frac{1}{T} \sum_{t=0}^{T-1} \frac{5K^2L^2\sigma^2\eta_t^3}{2}.
\end{eqnarray}
When $\eta_t \leq 1/\pth{2 \sqrt{30}KL}$, we have
\begin{equation*}
\eta_t - 30K^2L^2 \eta_t^3 \geq \eta_t  \pth{1-\sqrt{30}KL \eta_t} \geq \eta_{\min} - \sqrt{30}KL \eta_{\min}^2.
\end{equation*}
Then, Eqn.~\eqref{eqn:lsgd_5} can be further bounded as
\begin{eqnarray} \label{eqn:lsgd_6}
\frac{1}{T} \sum_{t=0}^{T-1} \Expect \norm{ \nabla f(x_{t}) }^2 &\leq& \frac{\frac{2}{K} \pth{f(x_0) - f^*} }{ T \pth{\eta_{\min} - \sqrt{30}KL \eta_{\min}^2} } +   \frac{5KL^2\sigma^2}{T \pth{\eta_{\min} - \sqrt{30}KL \eta_{\min}^2}} \sum_{t=0}^{T-1} \eta_t^3 \nonumber \\
&& + L \sigma^2 \frac{1}{T \pth{\eta_{\min} - \sqrt{30}KL \eta_{\min}^2}} \sum_{t=0}^{T-1} \frac{\eta_t^2}{n_t}.
\end{eqnarray}
Plugging in $\eta_t = \eta \sqrt{n_t/T}$, $\eta_{\min} = \eta \sqrt{n_{\min}/T}$, and using
\begin{equation*}
\sum_{t=0}^{T-1} \eta_t^3 \leq \frac{1}{\sqrt{T}} \eta^3 n_{\max}^{\frac{3}{2}}
\end{equation*}
lead to Eqn.~\eqref{eqn:nc_convg_lsgd}, and the proof is complete.
\end{Proof}

\end{document}